\documentclass[conference]{IEEEtran}

\usepackage{graphicx}
\usepackage{epstopdf} 
\usepackage{url}
\usepackage[usenames,dvipsnames]{color}
\usepackage{listings}
\definecolor{MyDarkGreen}{rgb}{0.0,0.4,0.0}

\ifCLASSINFOpdf
\else
\fi
\begin{document}
%
\title{Rethinking Electrostatic Solvers in Particle Simulations for the Exascale Era}

\author{\IEEEauthorblockN{Stefano Markidis \\  Rossen Apostolov \\  Erwin Laure}
\IEEEauthorblockA{PDC Center for High Performance Computing  \\
KTH Royal Institute of Technology \\
Stockholm, Sweden\\
Email: \{markidis, rossen, erwinl\}@pdc.kth.se}
\and
\IEEEauthorblockN{Giovanni Lapenta}
\IEEEauthorblockA{Center for Plasma Astrophysics\\
Wiskunde Department \\
KU Leuven \\
Leuven, Belgium\\
Email: giovanni.lapenta@wis.kuleuven.be}
}

%


\maketitle

\begin{abstract}
In preparation to the exascale era, an alternative approach to
calculate the electrostatic forces in Particle Mesh (PM) methods is
proposed. While the traditional techniques are based on the
calculation of the electrostatic potential by solving the Poisson
equation, in the new approach the electric field is calculated by
solving the Ampere's law. When the
Ampere's law is discretized explicitly in time, the electric field
values on the mesh are simply updated from the previous values. In
this way, the electrostatic solver becomes an embarrassingly parallel
problem, making the algorithm extremely scalable and suitable for
exascale computing platforms. An implementation of a one dimensional
PM code is presented to show that the proposed method produces correct
results, and it is a very promising algorithm for exascale PM
simulations.

\end{abstract}


%
\IEEEpeerreviewmaketitle

\section{Introduction}
Exascale supercomputers will deliver $10^{18}$ flop/s and will be likely composed of thousand million cores \cite{Kogge:2008}. The development of software for exascale computing platforms faces unprecedented challenges \cite{IESP_Roadmap}. The efficient exploitation of the exascale extreme parallelism is among the most important ones \cite{Kogge:2008}. Many traditional algorithms require by design implementations with local and global parallel communication that lead to a decrease of the code scalability. The electrostatic solvers in Particle Mesh (PM) methods are part of these algorithms, that require frequent and global communication, and are often cause of limited scalability of particle simulations. Scalability will become on exascale computing platforms even a more severe problem, that needs to be addressed to avoid the degradation of the performance.

It is a strong belief in High Performance Computing (HPC) community that not only new developments and enhancements of programming models, runtime systems, auto-tuners and compilers are required, but also the development of new algorithms and the reformulation of traditional methods are crucial \cite{Sarkar:2009} to exploit efficiently the exascale supercomputers. The design of new algorithms and the reformulation of well-known algorithms specifically for the exascale computing are topic of many recent investigations \cite{Keyes:2011}. These studies include communication avoiding iterative sparse solvers \cite{Demmel:2008}, task-based reformulation of linear algebra algorithms \cite{Song:2009}, synchronization reducing algorithms \cite{Agullo:2009}, and mixed precision calculations \cite{Baboulin:2009}. New algorithms for exascale particle simulations have been also proposed. For instance, event-driven simulations with asynchronous methodology \cite{Karimabadi:2005} and the gap-tooth methods are promising new algorithms \cite{Gear:2003} for exascale computing.

In this article, we propose a reformulation of the electrostatic solvers for particle simulations in oder to achieve scalability on future exascale supercomputers. Electrostatic solvers are the main computational engine of many application codes for bio-molecular, material science and plasma systems. For instance, the Particle-Particle-Particle-Mesh (P$^3$M) method \cite{Hockney:1981} for Molecular Dynamics (MD) simulations includes an electrostatic solver for calculating the long-range electrostatic forces. Among the most famous MD software packages, GROMACS \cite{Lindhal:2001} comprises a P$^3$M method and a PM electrostatic solver. Moreover, P$^3$M methods and electrostatic solvers are extensively used in material science studies\cite{Hockney:1981}.  Finally, the simulation of plasmas for fusion reactors, astrophysics, and space physics require electrostatic solvers also \cite{Markidis:2010}.  Because electrostatic solvers are present in many PM codes, the development of fast and scalable electrostatic solver is crucial for many science fields.

The proposed method is based on solving the explicit discretization of the Ampere's law (one of the four Maxwell's equations). This approach has been proposed previously by Markidis and Lapenta in Refs. \cite{Markidis:2011,Lapenta:2011} and by Chen {\em et al.} in Ref. \cite{Chen:2011} for carrying out plasma simulations. In both works the numerical scheme is based on the implicit discretization of the Ampere's law and particle equations of motion. These previous methods lead to the formulation of the first energy conserving PM methods. However, they require the solution of non-linear and linear solvers, making the algorithm not scalable on massively parallel supercomputer. Instead, it is here proposed to use the explicit discretization of the Ampere's law. The main advantage of this new formulation is that the calculation of the electric field becomes an embarrassingly parallel problem, not requiring parallel communication on supercomputers. 

The goal of this article is to present an electrostatic solver, extremely scalable and therefore suitable for extreme parallelism of exascale supercomputers. The article is organized as follows. In the second section, the traditional electrostatic solvers in PM methods are described. The proposed electrostatic solver, and in particular the governing equations, the algorithm, and numerical properties, and the suitability for exascale computing are presented in the third section. In the fourth section, the new electrostatic solver is verified against a standard benchmark of computational plasma physics to prove its correctness. The computational performance of the new method is presented in Section 5, and Section 6 concludes the article summarizing the main results.

\section{Electrostatic Solver in Particle Mesh Methods}
In PM electrostatic methods, the evolution of the system is determined by computing the trajectories of interacting particles via an electric field. The particle positions and velocities ($ \mathbf{x}_p$, $ \mathbf{v}_p$) are calculated by solving the equation of motion for each particle:
\begin{equation}
\label{eom11}
\left\{
\begin{array}{l}
d \mathbf{x}_p /d t =  \mathbf{v}_p \\ 
m_p d \mathbf{v}_p / dt  = q_p \mathbf{E}_p 
\end{array}
\right. ,
\end{equation}
where $\mathbf{E}_p$ is the electrostatic field acting on the particle $p$, and $q_p$ and $m_p$ are the particle charge and mass.
The most common and simple technique to discretize these equations is the leapfrog scheme, also known as Verlet algorithm \cite{Hairer:2006}:
\begin{equation}
\label{eom}
\left\{
\begin{array}{l}
\mathbf{x}_p^{n+1} = \mathbf{x}_p^n + \mathbf{v}_p^{n+1/2} \Delta t \\ 
\mathbf{v}_p^{n+3/2} = \mathbf{v}_p^{n+1/2} + \frac{q_p}{m_p}\mathbf{E}_p^{n+1} (\mathbf{x}_p^{n+1}) \Delta t
\end{array}
\right. ,
\end{equation}
where where $n$ is the time level. In the leap-frog method, particle positions are calculated at an integer time level while the velocities at a half-integer time level. Other forces might be present in the system, and their contributions should be included in the right hand of the second equation of the system above. The leapfrog scheme is a symplectic method, second order accurate in time and very simple to be implemented in computer codes. Other schemes can be used to increase the accuracy order and enhance numerical properties \cite{Hairer:2006}. 

In particle simulations, the main difficulty is to calculate the electric field $\mathbf{E}_p^{n+1}$ acting on the particles. It can be computed directly by adding the contributions from all the other particles, exerting a force on the particle. However, this method becomes unfeasible in practice when a large number of particles is present in the system. Instead, in the PM methods the electric field is calculated by introducing a mesh, and mapping the particle charge density $\rho_g$ to the mesh by using interpolation function $W(\mathbf{x}_g - \mathbf{x}_p)$:
\begin{equation}
\rho_g = \sum_p q_p W(\mathbf{x}_g - \mathbf{x}_p) / V_g,
\end{equation}
where $V_g$ is the volume of the cell $g$. Typically, the b-spline functions are chosen as interpolation functions. Once the charge density on the grid is known, it is possible to calculate the electrostatic field $\Phi$ by solving the Poisson equation:
\begin{equation}
\label{Poisson2}
\nabla^2 \Phi = - \rho/ \epsilon_0,
\end{equation}
where $\epsilon_0$ is the vacuum dielectric constant. After $\Phi$ is calculated, the electric field is computed as $\mathbf{E} = - \nabla \Phi$. The traditional electrostatic solvers all include the solution of the Poisson equation.
Finally, the electric field acting on the particle $p$ is calculated by interpolating electric fields on the mesh points back to the particles:
\begin{equation}
\mathbf{E}_p = \sum_p \mathbf{E}_g W(\mathbf{x}_g - \mathbf{x}_p) .
\end{equation}
In this article, it is proposed to replace the solution of the Poisson equation (Equation \ref{Poisson2}) with the calculation of Ampere's law.

\section{Exascale Ampere Electrostatic Solver}
In classical electromagnetism, the Ampere's law relates the magnetic field $\mathbf{B}$ to the electric current $\mathbf{J}$ and to the change in time of the electric field $\mathbf{E}$:
\begin{equation}
\nabla \times \mathbf{B} = \mu_0 \mathbf{J} + \mu_0 \epsilon_0 \frac{ \partial \mathbf{E}}{\partial t},
\end{equation}
where $\mu_0$ is the magnetic permittivity of vacuum.
In the electrostatic limit, the magnetic field $\mathbf{B}_0$ is fixed so that $ \mu_0 \mathbf{J}_0 =  \nabla \times \mathbf{B}_0 $. In this case, the Ampere's law reduces to:
\begin{equation}
\nabla \times \mathbf{B_0} =  \mu_0 \mathbf{J}_0   =  \mu_0 \mathbf{J} + \mu_0 \epsilon_0 \frac{ \partial \mathbf{E}}{\partial t},
\end{equation}
so that:
\begin{equation}
\displaystyle  \partial \mathbf{E} / \partial t =  (\mathbf{J_0} - \mathbf{J}) /  \epsilon_0 .
\label{Ampere1}
\end{equation}
In a system with no initial current $\mathbf{J}_0$, the Ampere's law simply becomes:
\begin{equation}
\displaystyle  \partial \mathbf{E} / \partial t = -  \mathbf{J} /  \epsilon_0 .
\label{Ampere1}
\end{equation}
If this equation is solved numerically, it can be simply discretized in time as follows:
\begin{equation}
(\mathbf{E}^{n+1} - \mathbf{E}^n) / \Delta t = - \mathbf{J}^{n+1/2}  /  \epsilon_0  .
\end{equation}
Alternative discretizations of equation \ref{Ampere1}  can be used to achieve higher order accuracies, but they are not here considered. 
The electric field can be calculated from the previous equation as:
\begin{equation}
\mathbf{E}^{n+1} = \mathbf{E}^n - \mathbf{J}^{n+1/2}  \Delta t  /  \epsilon_0 .
\label{solveE}
\end{equation}
Therefore, the electric field field can be simply updated using the current $\mathbf{J}$. The charge density $\rho$ is not calculated during the computational cycle, as in the traditional electrostatic solvers. If a charge unbalance, and consequently an electrostatic field, is present at the beginning of the simulation, the electric field need to be calculated consistently by solving the Poisson equation on the first computational cycle of the simulation. In addition, Equation \ref{solveE} does not require boundary conditions on the electric field. However, the boundary conditions should be enforced at the beginning of the simulation.

The current $\mathbf{J}^{n+1/2}_g$ is calculated by interpolation:
\begin{equation}
\label{Jmedio}
\mathbf{J}^{n+1/2}_g = \sum_p q_p \mathbf{v}^{n+1/2}_p W(\mathbf{x}_g - \mathbf{x}^{n+3/2}_p)/V_g .
\end{equation}
The interpolation is carried out using the particle position at time level $n+3/2$, that can be calculated as follows:
\begin{equation}
\mathbf{x}^{n+3/2} = \mathbf{x}^{n+1} + \mathbf{v}^{n+1/2}\Delta t /2 . 
\end{equation}

\begin{figure}[!t]
\centering
\includegraphics[width=3.0in]{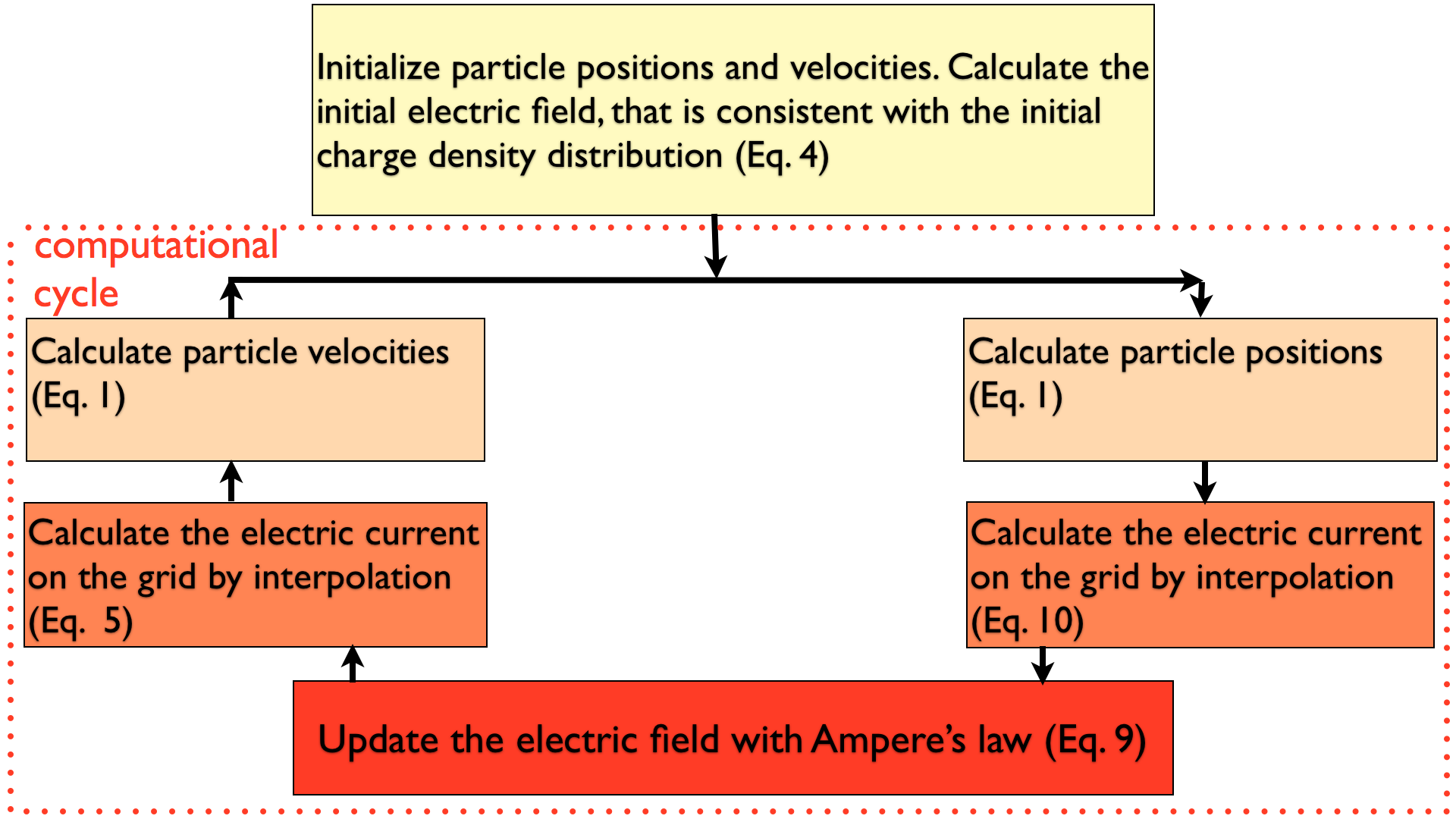}
\caption{Algorithm of the PM method with the Ampere electrostatic method. After the initialization, where a self-consistent electric field is calculated, the computational cycle is repeated many times.}
\label{Algo}
\end{figure}

In summary, the PM algorithm with the Ampere electrostatic solver is presented in Figure \ref{Algo}. After the particle positions and velocities are initialized, an electric field is calculated by solving the Poisson equation. The computational cycle is then repeated many times. Each cycle includes the update of particle positions, the calculation of the current on the grid points, the update of the electric field, the calculation of the electric field acting on the particles, and finally the update of particle velocities.

\subsection{Suitability for Exascale Computing}
The proposed Ampere electrostatic solver offers many advantages to the respect to the traditional approach. As shown before, the calculation of  the electric field is simply an update of the previous electric field values, and therefore it requires only $N_g$ operations. Instead, the solution of the Poisson equation requires the solution of sparse diagonal linear system or the computation of Fast Fourier Transform (FFT). Typically, these algorithms have $\mathcal{O}(N_g \log(N_g))$ computational cost, and only in particular cases they can achieve $\mathcal{O}(N_g )$ performance. 

The proposed solver is perfectly suited for exascale massively parallel computing. In fact, to update the values of the electric fields using Equation \ref{solveE} is an embarrassingly parallel problem. The electrostatic solver in the simple formulation does not require any parallel communication and therefore has ideal scalability. On the contrary, the traditional approach of solving the Poisson equation requires point-to-point and global parallel communication limiting already the scalability on current petascale computing platforms. 

\subsection{Interoperability with Traditional Electrostatic Solvers}
One of the main advantages of the Ampere's solver is that very small changes are required to include the proposed electrostatic solver in current production codes. Going from petascale to exascale, it is likely that an incremental approach will be chosen in the existing petascale applications, while disruptive algorithms requiring the use of new data structures will be avoided. The electrostatic Ampere's solver can be implemented using the existing data structures of the PM codes. The only modifications are to add a variable for the vector current $\mathbf{J}$ on the grid points, and to replace the charge deposition of the traditional approach with the current $\mathbf{J}$  deposition. 

\subsection{Towards Asynchronous and Adaptive Electrostatic Solvers}
It is clear from Equation \ref{solveE}, that the electric field depends on the value of the electric current. Spatial regions, characterized by strong currents, are subject to strong variations of the electric fields, while other regions with null or small currents have approximately unaltered electric field. Thus, a prompt update of the electric field is required only in regions with strong current. For instance, when an object is immersed in a plasma, the boundary layer between the object and the plasma is the spatial region with strong currents and therefore with strong variation of the electric field. Instead, far from the object, the currents vanish and therefore the electric field remains constant and the update of the electric field is not necessary. With the Ampere electrostatic solver it is possible to update the electric field only in regions with strong currents, while neglecting other regions with no currents. This allows user to adapt the time step locally in region with strong electric current, eliminating the need of dedicating computing resources to solve the electric field in spatial regions with null electric currents.

\subsection{Numerical Constraints Arising from the Other Maxwell's Equations}
The evolution of the electric and magnetic field ($\mathbf{B}$) is governed by the four Maxwell's equations:
\begin{equation}
\label{Maxwell}
\left\{
\begin{array}{l}
\displaystyle \nabla \cdot \mathbf{E} = \rho/\epsilon_0 \\
\displaystyle \nabla \cdot \mathbf{B} = 0 \\
\displaystyle \nabla \times \mathbf{E} = - \partial \mathbf{B} / \partial t  \\ 
\displaystyle  \nabla \times \mathbf{B} = \mu_0 \mathbf{J} + \mu_0 \epsilon_0 \partial \mathbf{E} / \partial t
\end{array}
\right. .
\end{equation}
In the electrostatic limit, the Maxwell's equations reduce to:
\begin{equation}
\label{MaxwellES}
\left\{
\begin{array}{l}
\displaystyle \nabla \cdot \mathbf{E} = \rho/\epsilon_0 \\
\displaystyle \nabla \times \mathbf{E} = 0 \\ 
\displaystyle  \partial \mathbf{E} / \partial t = (\mathbf{J}_0 -  \mathbf{J} )/  \epsilon_0 
\end{array}
\right. .
\end{equation}
This system comprises three equations that must be satisfied concurrently to describe correctly the evolution of the electric field. Therefore, the electric field calculated with Equation \ref{solveE} should satisfy additional two constraints: the electrostatic field must be consistent with the Gauss' ($\nabla \cdot \mathbf{E} = \rho/\epsilon_0$) and the Faraday's ($\nabla \times \mathbf{E} = 0$) laws. 

In fact, when the Poisson equation is solved in the traditional approach, a combination of Gauss' and Faraday's laws is computed. In fact, the electrostatic force is a conservative force, and therefore the electric field is irrotational (it satisfies the Faraday's law):
\begin{equation}
\nabla \times \mathbf{E} = 0 ,
\end{equation}
and therefore $\mathbf{E}$ can be written as a gradient of a scalar function $\Phi$:
\begin{equation}
\mathbf{E}  = - \nabla \Phi .
\end{equation}
Substituting this last expression into the Gauss' law, it is obtained:
\begin{equation}
\nabla \cdot \mathbf{E} = - \nabla \cdot \nabla \Phi = \rho/\epsilon_0 \Longrightarrow \nabla^2 \Phi = - \rho/\epsilon_0 .
\end{equation}

On the contrary, in the proposed new approach the electric field is advanced in time using the Ampere's law, and there is no guarantee that the Gauss' and Faraday's laws are satisfied also. These two laws can be regarded as conservation laws, as it is done for other macroscopic quantities that are not conserved in computer simulations, such as the total energy and momentum of the system. However, different techniques have been proposed to enforce$\nabla \cdot \mathbf{E} = \rho / \epsilon_0$ while solving the Ampere's law. Markidis and Lapenta suggested to use the method of pseudo-current, introduced by Marder \cite{Markidis:2011}. This technique reduces the numerical error but it does not eliminate it. It has been proved that a particular interpolation scheme can eliminate this error \cite{Chen:2011}.

\subsection{Energy and Momentum Conservation}
In the chosen discretization of the Ampere's law and particle equations of motion, both the total energy and momentum are not conserved. The traditional approach with Poisson electrostatic solver  leads to the exact conservation of the total momentum \cite{Birdsall:1985}, while the implicit formulation of the Ampere law and particle equation motion leads to the exact energy conservation \cite{Markidis:2011, Chen:2011}. In general, the variation of the momentum and energy depends on the time step, number of particles, and kind and order of the interpolation functions \cite{Hockney:1981}. A small time step, an increased number of particles and higher order interpolation functions lead to energy and momentum conservation at higher degree. The total energy and momentum of the system should be monitored and simulation should be disregarded when a in important violation (typically \textgreater   5\%) occurs. The disadvantage of the proposed solver is a higher variation of  the conserved quantities (total energy, momentum). An important violation of such conserved quantities might lead to numerical instabilities.

\section{Verification of the Ampere Electrostatic Solver}
The electrostatic Ampere solver has been implemented in a one dimensional Matlab/Octave PM code for plasma simulations. The code is presented in the Appendix of this article. The algorithm has been tested against a standard benchmark where the analytical calculation provides a result simulation can be compared with. A well-known benchmark, called {\em two stream instability}, has been used \cite{Birdsall:1985}. The instability grows exponentially in time according $\exp(\gamma t)$ law in the linear regime of the instability. After the linear phase, the non linear effects become important, and analytical theory and simulation can not be compared further. The analytical theory provides the growth rate ($\gamma$) of the instability for a given wavenumber $k = 2 \pi/\lambda$, where $\lambda$ is the wavelength of the spectral component. This exponential growth appears as a line in a semilogarithmic plot (red line in Figure 2).

\begin{figure}[!t]
\centering
\includegraphics[width=2.5in]{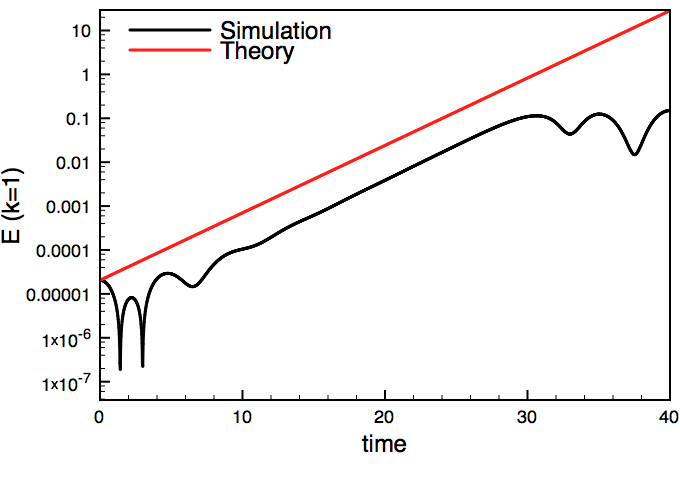}
\caption{The instability growth is calculated with analytical theory (red line) and simulated with the Ampere's electrostatic solver (black line). The semi-logarithmic plot of the spectral component $k = 1/ d_e$ of the electric field shows that the simulated instability has the same growth rate predicted by the theory (same slope of the curves). This comparison verifies the correctness of the Ampere electrostatic solver.}
\label{Comp_linear_theory}
\end{figure}

\begin{figure}[!t]
\centering
\includegraphics[width=2.5in]{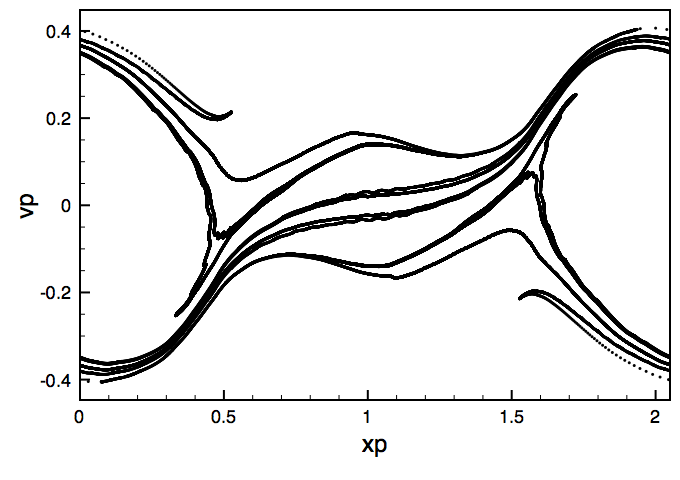}
\caption{Phase space plot of the system at time $\omega_{p} t = 40$. Each point represents one of the 100,000 particles with a given position a velocity. The instability leads to the formation of the characteristic electron {\em hole} in the phase space.}
\label{Phase_Space}
\end{figure}

In the {\em two stream instability} benchmark, two electron beams move in opposite directions with $ \pm 0.2 \ c$ speed, where $c$ is the speed of light in vacuum. A population of protons is present, but do not move and contribute only with a background charge density. The system is $2.05 c/\omega_p$ long and periodic. In this set-up, the growth rate $\gamma$ for $k=1/d_e$ ($d_e$ is the electron skin depth) in the linear regime can be calculated and it is equal to $0.35 \  \omega_p$, where $\omega_p$ is the plasma frequency. A PM mesh simulation with 100,000 particles, 256 grid points, a time step $dt = 0.025 \ \omega_p^{-1}$, 1600 computational cycles, has been carried out.  Figure \ref{Comp_linear_theory} shows a comparison of the instability growth calculated with the analytical theory (only valid in the linear regime of the instability) and with the Ampere electrostatic solver. In the linear regime of the instability, the slope of the curves, calculated with the simulation and linear theory matches, and therefore the electrostatic solver provides the correct results. Figure \ref{Phase_Space} shows the characteristic configuration of the phase space of the stream instability during the non linear regime \cite{Birdsall:1985}. As discussed previously, the proposed method introduces an error arising from not satisfying the Gauss' law. Because the simulation is one dimensional, the Faraday's law $\nabla \times \mathbf{E} = 0$ is trivially satisfied. Simulation in two and tree dimensions are necessary to assess the importance of errors arising from the violation of Faraday's law, and eventually devise numerical techniques to reduce or eliminate it. Figures 4 and 5 show that the total energy and momentum of the system are not conserved but vary in time. However, the degree to which Gauss' law is satisfied and total energy and momentum are conserved depends strongly on the simulation parameters. For instance, it is found that a decrease of the simulation time step reduces the error arising from the Gauss' law. This result is presented in Figure \ref{Gausslaw1}, where the two stream instability has been simulated with different time step values. Moreover it is found (but not shown here) that reducing the time step has a beneficial effect on the conservation of the total energy and momentum also.

\begin{figure}[!t]
\centering
\includegraphics[width=2.5in]{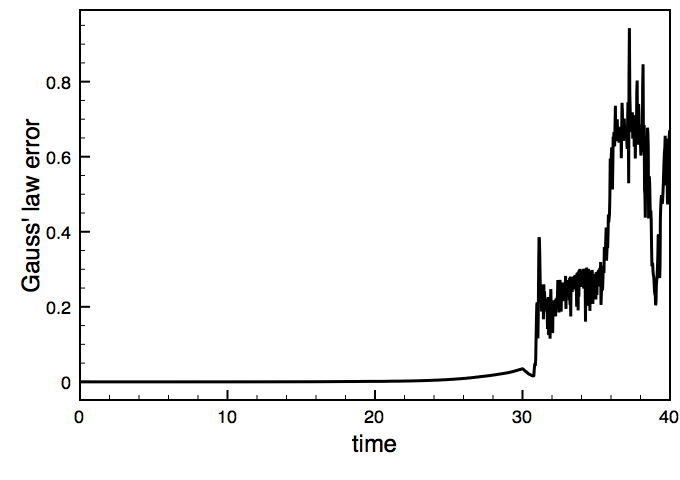}
\caption{History of $|\nabla \cdot \mathbf{E} - \rho| / \rho_{max}$, representing the degree the Gauss' law is not exactly satisfied.}
\label{Gausslaw}
\end{figure}

\begin{figure}[!t]
\centering
\includegraphics[width=2.5in]{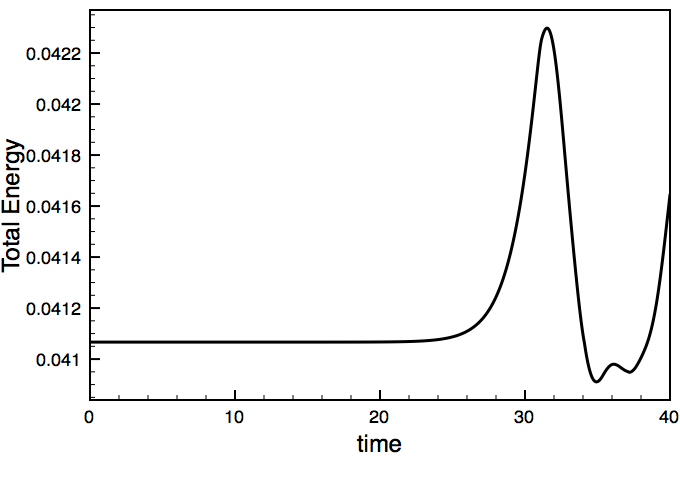}
\caption{History of the total energy of the system. The total energy of the system has maximum variation of the 2\%.}
\label{Total_Energy}
\end{figure}

\begin{figure}[!t]
\centering
\includegraphics[width=2.5in]{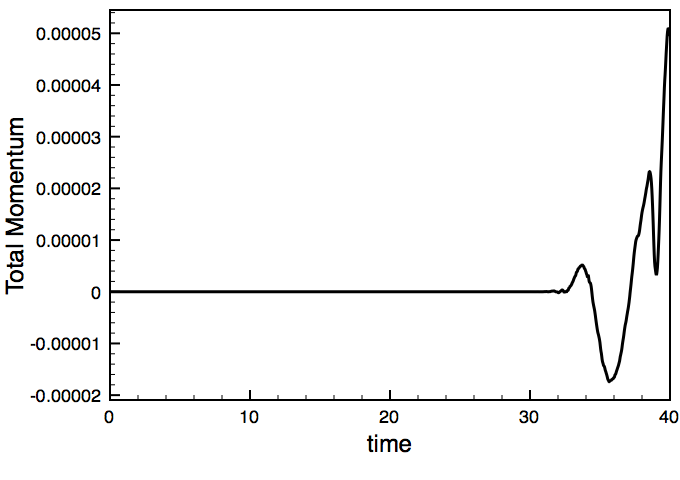}
\caption{History of the total momentum of the system. The total momentum is not exactly conserved.}
\label{Total_Momentum}
\end{figure}

\begin{figure}[!t]
\centering
\includegraphics[width=2.5in]{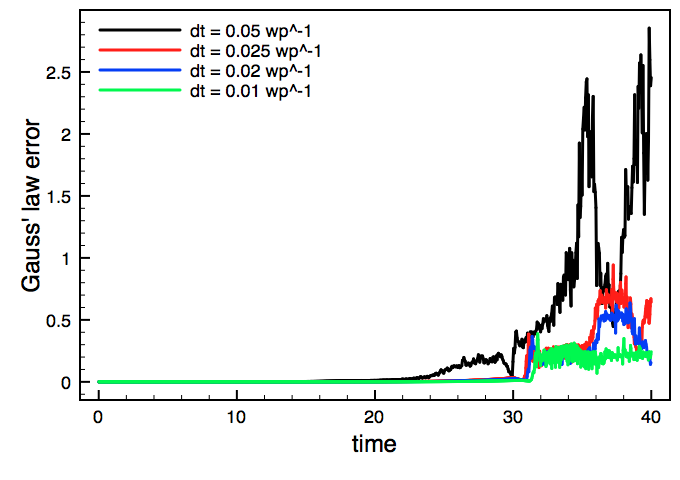}
\caption{History of $|\nabla \cdot \mathbf{E} - \rho| / \rho_{max}$, representing the degree the Gauss' law is not satisfied, using different values of the simulation time step $dt$. Decreasing the time step, the Gauss' law is satisfied at a higher degree. In addition, energy and momentum (not shown here) variations decrease with smaller time steps.}
\label{Gausslaw1}
\end{figure}

\section{Computational Performance}
\begin{figure}[!t]
\centering
\includegraphics[width=3.0in]{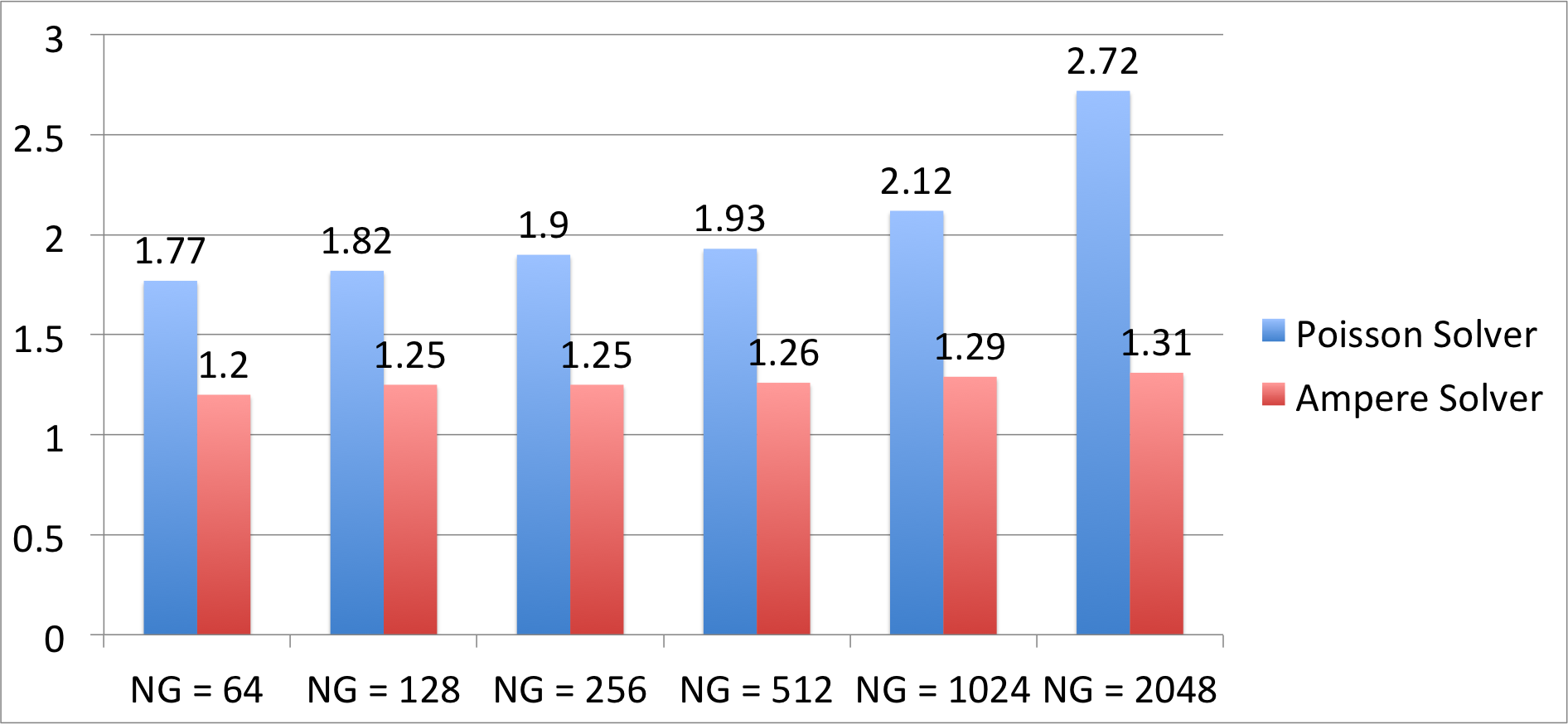}
\caption{Comparison of the execution time, expressed in seconds, of the proposed electrostatic solver (Ampere solver in red color) and the traditional approach (Poisson solver in blue color) with different number of grid points (NG). The computational time for the Poisson solver is higher than the one of the Ampere solver in all the cases. In addition the execution time of the Ampere's law increases increasing the number of grid points, while it remains approximately constant in the case of the Ampere solver.}
\label{ComputPerfo}
\end{figure}

The computational performance of the proposed algorithm and of the traditional approach, utilizing the Poisson equation, has been compared. The same simulation set-up presented in the verification study in Section 3 is chosen for the performance study. GNU Octave, version 3.2.3 version has been used on a 2.26 GHz Intel Core 2 Duo with 2 GB RAM memory on a 10.6.8 MAC OS X. The Poisson equation is solved by computing a tridiagonal linear system using Octave command, that chooses the most efficient solver for a given matrix. Figure \ref{ComputPerfo} shows a comparison of the execution time, expressed in seconds, of the electrostatic solver, based on the Poisson equation (in blue color) and on the Ampere's law (in red color) varying the number of grid points (NG). In all the cases, the Ampere's solver is faster than the traditional Poisson solver. In addition, the Ampere's solver has approximately constant execution time, while the Poisson solver execution time increases with the increase of the number of grid points.
The PM codes with Ampere's law does not allow to reuse the interpolation weight during the two interpolation stages. In fact, the current deposition is carried out using the particle positions at time level $n+3/2$:
\begin{equation}
\mathbf{J}^{n+1/2}_g = \sum_p q_p \mathbf{v}^{n+1/2}_p W(\mathbf{x}_g - \mathbf{x}^{n+3/2}_p)/V_g .
\label{Jmedio1}
\end{equation}
Instead, the electric field acting on the particle, used in Equation \ref{eom}, is calculated by using the particle positions at time level $n+1$:
\begin{equation}
\mathbf{E}^{n+1}_p = \sum_p \mathbf{E}^{n+1}_g W(\mathbf{x}_g - \mathbf{x}^{n+1}_p) .
\end{equation}
On the contrary, PM codes with traditional Poisson electrostatic solver requires only interpolations at time level $n+1$ allowing the reuse of the interpolation coefficients from  previous calculations.
\section{Conclusion}
An alternative approach to calculate the electrostatic fields in PM codes has been proposed. The method is based on solving the Ampere's law instead of the Poisson's equation. Because the proposed numerical scheme requires simply an update of the electric field using the values of the electric current, the new method results very scalable and suitable for the next exascale computing platforms. The computational cost of Ampere electrostatic solver is $\mathcal{O}(N_g)$ and does not require communication on parallel computers. On the contrary, traditional electrostatic solver, based on solving the Poisson equation, exhibits limited scalability because the algorithm requires parallel communication. The proposed electrostatic solver can be simply implemented in production codes without introducing disruptive changes, since it uses the same data structures of previous solvers. In addition, the proposed algorithm is suitable for the implementation of asynchronous particle simulation and adaptive technologies. Current work focuses on implementing an asynchronous PM method with the Ampere's electrostatic solver to remove the synchronizations barriers \cite{Karimabadi:2005}.  

A one dimensional implementation of a Matlab/Octave PM code has been developed to show that the Ampere electrostatic solver produces the correct results. The Gauss'  and Faraday's laws, the energy and momentum conservations might be violated at a certain degree when the Ampere's electrostatic solver is used. However, special numerical techniques can be devised to reduce and eliminate these errors.

\section*{Acknowledgment}
This research has received funding from the European Commission's FP7 Program inside the grant agreement CRESTA (cresta-project.eu).



%


\section*{Appendix}
The simple Octave/Matlab PM code with the Ampere's electrostatic solver is presented. 
\begin{lstlisting}
% Simulation parameters
L=2*pi/3.0600; % Simulation box length
DT= .025 % time step
NT= 1600 % number of computational cycles
NG= 256; % number of grid points
N=100000; % number of particles
WP=1; % plasma frequency
QM=-1; % mass ratio
V0=0.2; % beam velocity
VT=0.0; % thermal velocity
dx=L/NG; % grid spacing
XP1=1; V1=0.0; mode=1; % perturbation values
Q=WP^2/(QM*N/L); % particle charge
rho_back=-Q*N/L; % background density
xp=linspace(0,L-L/N,N)'; % initial particle positions
vp=VT*randn(N,1); % Maxwellian velocity distribution
pm=[1:N]'; pm=1-2*mod(pm,2);
vp=vp+pm.*V0; % 2 beams
% Perturbation
xp=xp+XP1*(L/N)*sin(2*pi*xp/L*mode);
% make Poisson tridiagonal matrix
p=1:N;p=[p p]; un=ones(NG-1,1);
Poisson=spdiags([un -2*un un],[-1 0 1],NG-1,NG-1);
% calculate the charge density rho from particle positions
g1=floor(xp/dx-.5)+1; g=[g1;g1+1];
fraz1=1-abs(xp/dx-g1+.5); fraz=[fraz1;1-fraz1];
out=(g<1);g(out)=g(out)+NG; out=(g>NG);g(out)=g(out)-NG;
mat=sparse(p,g,fraz,N,NG);
% calculate initial charge density
rho=full((Q/dx)*sum(mat))'+rho_back; 
% solve Poisson equation
Phi=Poisson\(-rho(1:NG-1)*dx^2);Phi=[Phi;0];
% Calculate E as  - gradient of Phi
Eg =([Phi(NG); Phi(1:NG-1)]-[Phi(2:NG);Phi(1)])/(2*dx);;
for it=1:NT
   xp=xp+vp*DT;  
   % apply periodic boundary conditions
   out=(xp<0); xp(out)=xp(out)+L;
   out=(xp>=L);xp(out)=xp(out)-L;
   % calculate the x at n+1/2 time level
   x_average = xp + vp*DT/2;
   g1=floor(x_average/dx-.5)+1; g=[g1;g1+1];
   fraz1=1-abs(x_average(1:N)/dx-g1+.5); fraz=[(fraz1);1-fraz1];	
   out=(g<1);g(out)=g(out) + NG; out=(g>NG);g(out)=g(out)- NG; 
   mat=sparse(p,g,fraz,N,NG);
   % calculate the J
   fraz=[(fraz1).*vp;(1-fraz1).*vp];	
   mat=sparse(p,g,fraz,N,NG);
   J = full((Q/dx)*sum(mat))'; 
   % calculate E with Ampere's law with 
   Eg = Eg - J*DT;
   % calculate Ep from xp^{n+1}
   g1=floor(xp/dx-.5)+1; g=[g1;g1+1];
   fraz1=1-abs(xp/dx-g1+.5); fraz=[fraz1;1-fraz1]	;
   out=(g<1);g(out)=g(out)+NG; out=(g>NG);g(out)=g(out)-NG;
   mat=sparse(p,g,fraz,N,NG);
   % update particle velocity
   vp=vp+mat*QM*Eg*DT;
end
\end{lstlisting}
\end{document}